 \documentclass[aps,pra,preprint,groupedaddress,amsmath,amssymb]{revtex4}
\usepackage{graphicx}
\usepackage{amsmath}
\usepackage{amssymb}
\usepackage{mathrsfs}
\usepackage{amsfonts}
\usepackage{dcolumn}
\usepackage{bm}

\newcommand{\vvr}{\mathbf{r}}

\newcommand{\vk}{\mathbf{k}}

\newcommand{\vE}{\mathbf{E}}

\newcommand{\hvk}{\hat{\mathbf{k}}}
\newcommand{\hve}{\hat{\mathbf{e}}}

\newcommand{\hvx}{\hat{\mathbf{x}}}
\newcommand{\hvy}{\hat{\mathbf{y}}}
\newcommand{\hvz}{\hat{\mathbf{z}}}
\providecommand{\abs}[1]{\lvert#1\rvert}

\newcommand{\di}{\mathrm{d}}

\begin{document}
\title{Theory of angular Goos-H\"{a}nchen shift near Brewster incidence}
\author{A. Aiello$^{1,2*}$}
\author{J. P. Woerdman$^{2}$}
\affiliation{${^1}$Max Planck Institute for the Science of Light, G\"{u}nter-Scharowsky-Str. 1/Bau 24, 91058 Erlangen, Germany}
\affiliation{${^2}$Huygens Laboratory, Leiden University
P.O.\ Box 9504, 2300 RA Leiden, The Netherlands}
\affiliation{$^*$Corresponding author: Andrea.Aiello@mpl.mpg.de}
\begin{abstract}
We present here a compactly formulated application of the previously posted general formalism of the reflection of Gaussian beams at a dielectric interface ({arXiv:0710.1643v2 [physics.optics]}). Specifically, we calculate the  Goos-H\"{a}nchen shift near Brewster incidence, for an air-glass plane interface.
\end{abstract}
\maketitle
\section{Introduction}
When a beam of light impinges upon a plane interface separating
two transparent media, it produces reflected
and  transmitted beams.
In 1815 the Scottish physicist David Brewster discovered the total polarization
of the reflected beam at the angle $\theta_B$  since named after
him \cite{Brewster1815}. From his  observations he was also able to empirically determine the celebrated equation, known as Brewster's law, $\tan \theta_B = n_1/n_2$, where  $n_1$ and $n_2$ are the respective
refractive indices of the two media.

In this work we calculate the  Goos-H\"{a}nchen shift occurring near Brewster incidence at an air-glass plane interface, for an incident Gaussian beam.
\section{Theory}
Consider a monochromatic beam of light incident upon a plane interface  that separates two homogeneous and isotropic media. The first medium, say air, has refractive index $n_<$ and the second medium, say glass, has refractive index $n_{>}$. With $n = \frac{n_{>}}{n_{<}}$ we denote the ratio between the two refractive indices. Here $n$ can be either a real or a complex number, in the latter case at least one of the two media exhibits absorption. Without lack of generality, we assume that the beam meets the interface coming from the air side.
%
Thus, it will be convenient to take the axis $z$ of the laboratory Cartesian frame $K = (O,x,y,z)$ normal to the  interface and directed from the air to the glass. Moreover, we choose the origin $O$ in a manner that the plane interface has  equation $z=0$.
The air-glass interface, the incident and the reflected beams are pictorially illustrated in Fig. 1.
%

  \begin{flushleft}
  \begin{figure}[!ht]
  \centerline{\includegraphics[width=12truecm]{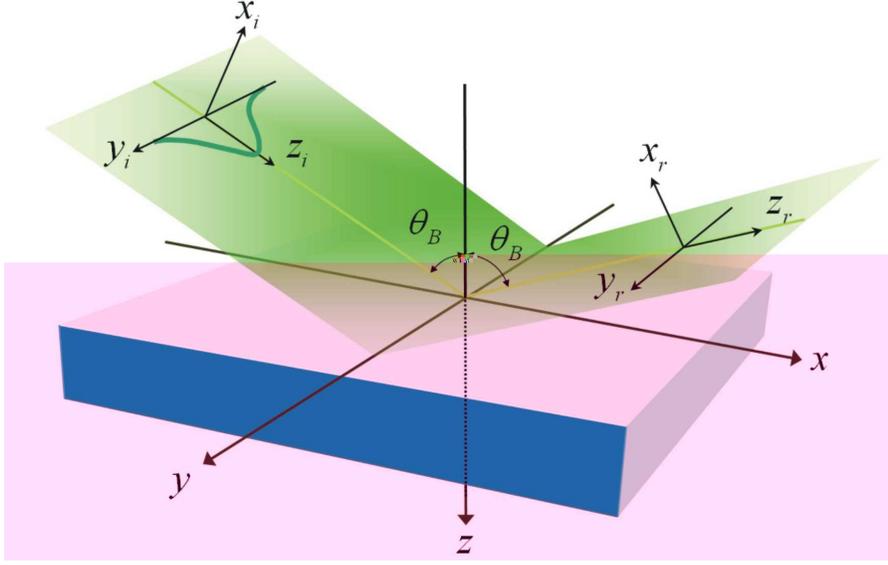}}
  \caption{(Color online) Geometry of beam reflection at the air-medium interface. $\theta_B$ is the Brewster angle.}
  \end{figure}
  \end{flushleft}
%
%
In addition to the laboratory frame,  we use a Cartesian frame $K_i=(O,x_i, y_i, z_i)$ attached to the incident beam and another one $K_r=(O,x_r, y_r, z_r)$ attached to the reflected beam.
Let $\mathbf{k}_0 = k_0 \hat{\mathbf{z}}_i$ and  $\mathbf{k}$ denote the central and noncentral wave vectors of the incident beam, respectively, with $\abs{\mathbf{k}} = \abs{\mathbf{k}_0} = k_0$. We choose the laboratory frame $K$ in such a way that $\hvz_i =  \hvx \sin \theta + \hvz \cos \theta$. In this manuscript with either $\hat{\mathbf{u}}$ or $\hat{\mathbf{u}}_{\,a}$  we denote a real unit vector directed along the Cartesian frame axis $u$, where $u \in \{ x,y,z\}$, and $ a \in \{ i,r\}$.

The electric field of the incident beam at the air side of the interface $(z<0)$,
can be written in the angular spectrum representation \cite{MandelBook} as
\begin{align}\label{eq10}
\mathbf{E}^I (\mathbf{r}) &  = \sum_{\lambda=1}^2 \iint\limits_{-\infty}^{\hphantom{xxii}\infty}  \hve_\lambda(U,V,\theta) E_\lambda(U,V) e^{i \left(  U X_i + V Y_i + W Z_i\right)} \; \di U \, \di V,
\end{align}
where $U = \vk \cdot \hvx_i /k_0 $, $V = \vk \cdot \hvy_i /k_0 $, and $W = (1 - U^2 - V^2)^{1/2}$. Moreover, we have defined $X_a = k_0 x_a$, $Y_a = k_0 z_a$, $Z_a = k_0 z_a$, with $a \in \{ i,r\}$, and $\vvr$ so .
The polarization unit basis vectors $\{ \hve_\lambda\}_{\lambda\in \{ 1,2\}}$  have been chosen as
\begin{align}\label{pra20}
\hve_1 =  \frac{\hve_2 \times{\mathbf{k}}}{\abs{\hve_2 \times{\mathbf{k}}}}, \qquad \hve_2 = \frac{\hat{\mathbf{z}} \times {\mathbf{k}}}{\abs{ \hat{\mathbf{z}} \times {\mathbf{k}} }},
\end{align}
where the symbol ``$\times$'' denotes the standard vector product in $\mathbb{R}^3$.
 Here $\hat{\mathbf{z}}$ is a real unit vector directed along the laboratory axis $z$ and, by definition,
\begin{align}
{\mathbf{k}} & = k_0 \left( U \hvx_i + V \hvy_i + W \hvz_i \right) \nonumber \\
 & =   \hvx k_x +  \hvy  {k_y} +  \hvz \sqrt{k_0^2 - k_x^2 - k_y^2}. \label{hk}
\end{align}
By substituting Eq. (\ref{hk}) into Eq. (\ref{pra20})  we obtain
\begin{align}
\hat{\mathbf{e}}_1(U,V, \theta) & = \frac{\left[U W c + \left( W^2 + V^2 \right) s \right] \hvx_i + V \left( W c - U s \right) \, \hvy_i
- \left[\left( U^2 + V^2 \right) c + U W s  \right] \, \hvz_i}{\left[ V^2 + \left(U c + W s \right)^2\right]^{1/2}}, \label{e1} \\
\hat{\mathbf{e}}_2(U,V, \theta) & = \frac{-{V c}\, \hvx_i + \left( U c + W s \right) \, \hvy_i -V s \, \hvz_i}{\left[ V^2 + \left(U c + W s \right)^2\right]^{1/2}}, \label{e2}
\end{align}
where we used the shorthand $c = \cos \theta$ and $s = \sin \theta$, and $\theta$ is the \emph{central} angle of incidence defined as: $\theta = \arccos (\hvz_i \cdot \hvz)$. For well collimated paraxial beams ($U^2 + V^2 \ll 1$) the expressions  above reduce to
\begin{align}
\hat{\mathbf{e}}_1 & \simeq { \hvx_i + \hvy_i \, V \cot \theta  - \hvz_i  \, U}, \label{e1b} \\
\hat{\mathbf{e}}_2 & \simeq {- \hvx_i \, V \cot \theta + \hvy_i -\hvz_i \, V }, \label{e2b} \\
\hat{\mathbf{e}}_3 & \simeq { \hvx_i \, U + \hvy_i V +\hvz_i  }, \label{e3b}
\end{align}
where $\hat{\mathbf{e}}_3 \equiv \vk/k_0 \equiv \hat{\vk}$.
From Eq. (\ref{pra20}) it follows that $\hve_1$ lies in the plane containing both the wave vector $\mathbf{k}$ and $\hat{\mathbf{z}}$, usually denoted as the plane of incidence with respect to $\mathbf{k}$, while $\hve_2$ is orthogonal to such a plane. Both $\hve_1$ and $\hve_2$ are orthogonal to $\mathbf{k}$ by definition, and $\{ \hve_1,\hve_2, \hvk \}$ form a complete, orthogonal basis in $\mathbb{R}^3$.
 Conventionally,  a plane wave whose \emph{electric} field is parallel to either $\hve_1$ or $\hve_2$, is referred to as either  a \textsl{TM} or a \textsl{TE}  wave, respectively. The symbols  \textsl{P} for  \textsl{TM} and \textsl{S} for \textsl{TE}, are also widely used.
In Eq. (\ref{eq10})  the functions   $E_\lambda(U,V)$ determine the shape and the polarization of the beam. These amplitudes are always expressible as
$E_\lambda(U,V) = A(U,V) \, \alpha_\lambda(U,V)$, where
 $A(U,V)$ and $\alpha_\lambda(U,V)$ are the scalar and the vector spectral amplitudes of the field, respectively.  The first determines the \emph{spatial} characteristics of the beam, while the second sets the \emph{polarization} of the beam \cite{AielloOL08}.
Here we consider a collimated, monochromatic beam, with a Gaussian spectral amplitude
\begin{align}\label{pra30}
A(U,V) =   \exp\left( - \frac{U^2 + V^2}{\theta_0^2}\right) \exp  \left( i W D \right),
\end{align}
where $\theta_0 = 2/(k_0 w_0)$ is the angular spread of the incident beam with a minimum spot size (waist) $w_0$ located at $Z_i = -D$ \cite{MandelBook}. In order to determine the vector spectral amplitudes $\alpha_\lambda(U,V)$ of the incident beam, we assume that the beam has passed across a polarizer plate perpendicular to the central wave vector $k_0 \hvz_i$ of the beam. Let $\hat{\mathbf{f}} = f_P \hvx_i + f_S \hvy_i$ denotes a complex-valued unit vector that represents the orientation of the polarizer, with $\abs{f_P}^2 + \abs{f_P}^2 =1$. Then, we determine the amplitudes $\alpha_\lambda(U,V)$ by imposing
\begin{align}\label{pra40}
\sum_{\lambda=1}^2 \hve_\lambda(U,V,\theta) \alpha_\lambda(U,V) = \hat{\mathbf{f}} - \hvk \bigl( \hvk \cdot \hat{\mathbf{f}} \bigr),
\end{align}
where we used the  polarizer representation given in Ref. \cite{FainmanANDShamir}. Since the completeness of the basis $\{ \hve_1,\hve_2, \hvk \}$ implies, for any  vector $\mathbf{v}$, the validity of the following relation
\begin{align}\label{pra50}
 \hve_1 ( \hve_1 \cdot \mathbf{v} ) + \hve_2  ( \hve_2 \cdot \mathbf{v} ) = \mathbf{v}  - \hvk  ( \hvk  \cdot \mathbf{v} ),
\end{align}
then from Eq. (\ref{pra40}) it immediately follows that
\begin{align}\label{pra60}
\alpha_\lambda(U,V) =  \hve_\lambda(U,V,\theta) \cdot \hat{\mathbf{f}}.
\end{align}
Thus, from Eqs. (\ref{e1}-\ref{e2}) and (\ref{pra60}) we obtain
\begin{align}
\alpha_1(U,V) & = \frac{W \left( U f_P + V f_S \right) \cos \theta - \left[ U V f_S - \left( W^2 + V^2\right) f_P   \right]\sin \theta }{\left[ V^2 + \left(U \cos \theta + W \sin \theta \right)^2\right]^{1/2}}, \label{a1}  \\
\alpha_2(U,V) & = \frac{ \left( U f_S - V f_P \right) \cos \theta + W f_S  \sin \theta  }{\left[ V^2 + \left(U \cos \theta + W \sin \theta \right)^2\right]^{1/2}},   \label{a2}
\end{align}
that reduce, for paraxial beams, to
\begin{align}
\alpha_1(U,V) & \simeq f_P + f_S V \cot \theta,  \label{a1b} \\
\alpha_2(U,V) & \simeq f_S - f_P V \cot \theta.  \label{a2b}
\end{align}
When the beam is reflected at the interface,  each plane wave  mode function
\begin{align}\label{eq100}
 \hve_\lambda(U,V,\theta)  e^{i \left(  U X_i + V Y_i + W Z_i\right)} =  \hve_\lambda(\vk)  e^{i \vk \cdot \vvr }, 
\end{align}
changes according to
\begin{align}
\hve_\lambda(\vk)  e^{i \vk \cdot \vvr } \mapsto  r_\lambda(\mathbf{k}) \hve_\lambda(\widetilde{\vk})  e^{i \widetilde{\vk} \cdot \vvr }, \label{change}
\end{align}
 where $r_1(\mathbf{k})$ and $r_2(\mathbf{k})$ are the  Fresnel reflection amplitudes for \textsl{TM} and \textsl{TE} waves, respectively \cite{BandWBook},
\begin{align}\label{eq110}
r_1(\mathbf{k}) = \displaystyle{\frac{n^2 k_z - k_z^t}{n^2 k_z + k_z^t}}, \qquad   \qquad r_2(\mathbf{k}) = \displaystyle{\frac{ k_z - k_z^t}{ k_z + k_z^t}},
\end{align}
and $ \widetilde{\mathbf{k}}= \mathbf{k} - 2 \, \hat{\mathbf{z}}\left(\hat{\mathbf{z}} \cdot \mathbf{k}  \right)$ is set by the law of specular reflection  \cite{Gragg}, while the unit vectors $\hve_\lambda(\widetilde{\vk})$ are defined as in Eq. (\ref{pra20}) with $\vk \mapsto \widetilde{\vk}$.
In Eqs. (\ref{eq110}) $k_z^t$ is the $z$-component of the wave vector inside the glass, namely
\begin{align}\label{eq120}
k_z^t =  \left(n^2 k_0^2 -k_x^2 -k_y^2\right)^{1/2} .
\end{align}
It is worth noting that here $k_x,k_y,k_z$ are the Cartesian components of the wave vector $\vk$ with respect to the laboratory frame $K$, while in Eq. (\ref{eq10})  the integrations are performed with respect to the variables $U$ and $V$ which are the transverse Cartesian components of the wave vector $\vk$ with respect to the incident-beam frame $K_i$. Therefore, it will be useful to express $r_1(\mathbf{k})$ and $r_2(\mathbf{k})$ in terms of  $U$ and $V$. From Eq. (\ref{hk}) it straightforwardly follows that
\begin{align}
r_1(U,V) & = \, \frac{n^2 \left(W \cos \theta - U \sin \theta \right) - \left[
n^2 - V^2 - \left(W \sin \theta + U \cos \theta \right)^2\right]^{1/2}}{n^2 \left(W \cos \theta - U \sin \theta \right) + \left[
n^2 - V^2 - \left(W \sin \theta + U \cos \theta \right)^2\right]^{1/2}}, \label{r1} \\
r_2(U,V) & = \, \frac{W \cos \theta - U \sin \theta - \left[
n^2 - V^2 - \left(W \sin \theta + U \cos \theta \right)^2\right]^{1/2}}{W \cos \theta - U \sin \theta  + \left[
n^2 - V^2 - \left(W \sin \theta + U \cos \theta \right)^2\right]^{1/2}}. \label{r2}
\end{align}
It is easy to check that $r_1(U,V)$ and $r_2(U,V)$ reduce to the ordinary Fresnel coefficients for $U=0$ and $V=0$:
\begin{align}
r_1^0(\theta) \equiv r_1(0,0) & = \, \frac{n^2  \cos \theta  - \left(
n^2 -\sin^2 \theta \right)^{1/2}}{n^2  \cos \theta  + \left(
n^2 -\sin^2 \theta \right)^{1/2}}, \label{r1b} \\
r_2^0(\theta)  \equiv r_2(0,0) & = \, \frac{ \cos \theta  - \left(
n^2 -\sin^2 \theta \right)^{1/2}}{  \cos \theta  + \left(
n^2 -\sin^2 \theta \right)^{1/2}}. \label{r2b}
\end{align}
In the remaining of this manuscript, we shall often benefit from the following relations satisfied by the reflection coefficients defined above:
\begin{align}
\left. \frac{\partial \, r_\lambda}{\partial U}\right|_{U=0,\, V=0} & = \frac{\partial \, r_\lambda^0}{\partial \theta}, \qquad
\left. \frac{\partial \, r_\lambda}{\partial V}\right|_{U=0,\, V=0}  = 0, \label{rc}
\end{align}
and
\begin{align}
\left. \frac{\partial^{\,2} \, r_\lambda}{\partial U^2}\right|_{U=0,\, V=0} & = \frac{\partial^2 \, r_\lambda^0}{\partial \theta^2}, \qquad
\left. \frac{\partial^{\,2} \, r_\lambda}{\partial V^2}\right|_{U=0,\, V=0}  = \cot \theta \frac{\partial \, r_\lambda^0}{\partial \theta}, \qquad \left. \frac{\partial^{\,2} \, r_\lambda}{\partial U \partial V}\right|_{U=0,\, V=0}  = 0,\label{rc2}
\end{align}
where $\lambda \in \{1,2\}$.
From Eq. (\ref{change}) it follows that, after reflection, the electric field of the  beam can be  written as
\begin{align}\label{Eref}
\mathbf{E}^I (\mathbf{r} )  \mapsto \mathbf{E}^R (\mathbf{r})
 &  = \sum_{\lambda=1}^2 \iint\limits_{-\infty}^{\hphantom{xxii}\infty}  \hve_\lambda(\widetilde{\vk}) r_\lambda(U,V) E_\lambda(U,V) e^{i \left( - U X_r + V Y_r + W Z_r\right)} \; \di U \, \di V,
\end{align}
where $\hve_\lambda(\widetilde{\vk})= \hve_\lambda(-U,V,\pi -\theta)$, namely:  
\begin{align}
\hve_1(\widetilde{\vk}) & = \frac{\left[U W c + \left( W^2 + V^2 \right) s \right] \hvx_r - V \left( W c - U s \right) \, \hvy_r
+ \left[\left( U^2 + V^2 \right) c + U W s  \right] \, \hvz_r}{\left[ V^2 + \left(U c + W s \right)^2\right]^{1/2}}, \label{g1} \\
\hve_2(\widetilde{\vk}) & = \frac{{V c}\, \hvx_r + \left( U c + W s \right) \, \hvy_r -V s \, \hvz_r}{\left[ V^2 + \left(U c + W s \right)^2\right]^{1/2}}, \label{g2}
\end{align}
where we used again the shorthand $c = \cos \theta$ and $s = \sin \theta$. In Eq. (\ref{Eref}) we have exploited the fact that by definition
\begin{align}
\widetilde{\vk} \cdot \vvr =  k_x x + k_y y - k_z z =  - U X_r + V Y_r + W Z_r,
\end{align}
where the latter equality is written in terms of the Cartesian coordinates of the position vector $\vvr$ with respect to the reflected beam reference frame $K_r$.

If the air-glass interface would behave as and ideal reflecting surface characterized by wave vector-independent reflection amplitudes  $r_1(\mathbf{k}) = 1$ and $r_2(\mathbf{k}) = -1$, then the reflected beam were just the mirror-image of the incident one \cite{Aiello}.  However, in the real world, as a result of the polarization and wave vector dependence of the Fresnel amplitudes
%
%
%
%
%
$r_\lambda(\mathbf{k})$, non-specular reflection phenomena occur,
the most prominent of which are the so-called
Goos-H\"{a}nchen  (GH) \cite{GH} and Imbert-Fedorov (IF) \cite{CandI} shifts that amount, respectively,  to a longitudinal and a transverse displacement  of the reflected beam with respect to the mirror-image of the incident one.
Such displacements can be assessed by measuring the position of the center of the reflected beam with a quadrant detector centered at $x_r=0, \, y_r=0$ along the reference axis $z_r$ attached to the reflected central wave vector $\widetilde{\mathbf{k}}_0 = k_0 \hat{\mathbf{z}}_r$. A quadrant detector has four
sensitive areas each delivering a photocurrent when illuminated. The difference between these photocurrents is proportional to the displacement of the barycenter of the beam intensity $I(X_r,Y_r,Z_r)$ relative to center of the detector. In other words, this displacement is proportional to the first order moments $\langle \mathbf{X} \rangle = \langle X_r  \rangle \hat{\mathbf{x}}_r + \langle Y_r  \rangle \hat{\mathbf{y}}_r $ of the intensity distribution function  of the beam \cite{Porras97}:
\begin{align}\label{Xmedio}
\langle \mathbf{X} \rangle =  \frac{\displaystyle{\iint\limits_{-\infty}^{\hphantom{xxii}\infty}} \mathbf{X} \, I(X_r,Y_r,Z_r) \text{d}X_r \text{d}Y_r}{\displaystyle{ \iint\limits_{-\infty}^{\hphantom{xxii}\infty} I(X_r,Y_r,Z_r) \text{d}X_r \text{d}Y_r}}.
\end{align}
In order to evaluate $\langle \mathbf{X} \rangle$ we need to know the intensity $I(X_r,Y_r,Z_r)$ that, apart from an irrelevant proportionality factor, can be defined as
\begin{align}\label{intensity}
I(X_r,Y_r,Z_r) \equiv \abs{\vE^R(X_r,Y_r,Z_r)}^2.
\end{align}
Thus, we must calculate the double integral in Eq. (\ref{Eref}). To this end, we exploit the fact that for a well collimated beam $\theta_0 \ll 1$, and that Eq. (\ref{pra30}) implies that $A(U,V) \simeq 0$ outside the paraxial domain $\mathcal{P}= \{U, V: U^2 + V^2 \ll 1 \} $.
In this domain
\begin{align}\label{W}
W = \left(1 -U^2  -V^2 \right)^{1/2} \simeq 1 - \frac{U^2 + V^2}{2},
\end{align}
and we can rewrite Eq. (\ref{Eref}) as
\begin{align}\label{Eref2}
 \mathbf{E}^R (\mathbf{r})
 &  = e^{i (Z_r + D)}\iint\limits_{-\infty}^{\hphantom{xxii}\infty}   \bm{\mathcal{E}} (U,V)e^{- \frac{U^2 + V^2}{2}\left[\Lambda +  i(Z_r + D) \right]} e^{i \left( - U X_r + V Y_r \right)} \; \di U \, \di V,
\end{align}
where we have defined $\Lambda = 2/\theta_0^2 = k_0 L$, with $L$ equal to the Raleigh range of the beam \cite{MandelBook}. Equation (\ref{Eref2}) is still exact, and it defines $ \bm{\mathcal{E}} (U,V)$ as
\begin{align}\label{vF}
 \bm{\mathcal{E}} (U,V)
 &  =  \displaystyle{\sum_{\lambda=1}^2   \hve_\lambda(\widetilde{\vk}) r_\lambda(U,V) \alpha_\lambda(U,V) e^{i  (Z_r + D)\left[ W - \left( 1 - \frac{U^2 + V^2}{2}\right)\right]}},
\end{align}
which can be evaluated within the paraxial domain $\mathcal{P}$ via a Taylor expansion of the form
\begin{align}\label{vFSeries}
 \bm{\mathcal{E}} (U,V)
 &  \simeq   \bm{\mathcal{E}} (0,0) + \left(U  \bm{\mathcal{E}}_U  +V  \bm{\mathcal{E}}_V  \right) + \frac{1}{2}\left(U^2  \bm{\mathcal{E}}_{UU} + 2 U V  \bm{\mathcal{E}}_{UV}  +V^2 \bm{\mathcal{E}}_{VV}  \right) + \ldots,
\end{align}
where we used the obvious notation $ \bm{\mathcal{E}}_U  = \left.\partial  \bm{\mathcal{E}}(U,V)/ \partial U \right|_{U=0 ,V=0}$, and so on. Usually, to calculate both GH and IF shifts, first order Taylor expansions is enough. However, as we shall see soon, at Brewster incidence  it becomes necessary to keep second order terms to avoid divergences in the expressions of the shifts.
Substitution of Eq. (\ref{vFSeries}) in Eq. (\ref{Eref2}) permits the analytical evaluation of the Gaussian integrals; this  leads to the following expression for the electric field of the reflected beam:
\begin{align}\label{vEref}
 \mathbf{E}^R (\mathbf{r})
 &  \simeq \psi(\vvr) \bigl( \hvx_r E_{x_r}^R +  \hvy_r E_{y_r}^R+\hvz_r E_{z_r}^R \bigr),
\end{align}
where
\begin{align}\label{ante}
\psi(\vvr)
 & = \frac{1}{\displaystyle{Z- i \Lambda}} \exp\left({\frac{\displaystyle{i}}{\displaystyle{2}} \frac{\displaystyle{X^2 + Y^2}}{\displaystyle{Z- i \Lambda}}}\right),
\end{align}
is the scalar amplitude of a fundamental Gaussian beam, and
\begin{align}
 E_{x_r}^R = & \,   f_P r_P+  f_S \frac{Y \left(r_P+r_S\right) \cot \theta }{Z-i \Lambda }- f_P  \frac{X r_P'}{Z-i \Lambda } \nonumber \\
& \, +  f_P\frac{X^2 \left(r_P'' -2 r_P \right)}{2 (Z-i \Lambda )^2}+ f_P\frac{Y^2  \left[ r_P'-2 \left(r_P+r_S\right) \cot \theta  \right] \cot \theta}{2 (Z-i \Lambda )^2} \nonumber  \\
& \, -i f_P \frac{  r_P''+2 r_S-2 \left(r_P+r_S\right)\csc^2 \theta +  r_P' \cot \theta  }{2 (Z-i \Lambda )}\nonumber  \\
& \,+ f_S\frac{X Y \left[2 r_P+r_S+r_S \cos(2 \theta) -\left(  r_P'+  r_S'\right)\sin(2 \theta )\right] \csc^2 \theta }{2 (Z-i \Lambda )^2}, \label{vErefx}\\
 %
 %
 %
 E_{y_r}^R  =  & \,f_S r_S- f_P\frac{Y \left(r_P+r_S\right) \cot \theta}{Z-i \Lambda }-f_S  \frac{X r_S'}{Z-i \Lambda }\nonumber  \\
& \,+ f_S\frac{X^2 r_S''}{2 (Z-i \Lambda )^2}+ f_S\frac{Y^2 \left[2 r_P-2 \left(r_P+r_S\right)\csc^2 \theta +  r_S' \cot \theta  \right]}{2 (Z-i \Lambda )^2} \nonumber  \\
& \,- i f_S \frac{r_S''+2 r_P-2\left(r_P+r_S\right)\csc^2 \theta +  r_S'\cot \theta }{2 (Z-i \Lambda )} \nonumber \\
& \,- f_P \frac{X Y \left[2 r_P+r_S+r_S\cos(2 \theta) -\left(   r_P'+   r_S'\right)\sin(2 \theta) \right]  \csc^2 \theta}{2 (Z-i \Lambda )^2}, \label{vErefy}\\
 E_{z_r}^R  =  & \,- f_P \frac{Xr_P}{Z-i \Lambda }- f_S \frac{Y r_S}{Z-i \Lambda } \nonumber \\
& \,+ f_P \frac{Y^2\left(r_P+r_S\right) \cot \theta}{(Z-i \Lambda )^2}+ f_P\frac{X^2    r_P'}{(Z-i \Lambda )^2}+ f_S \frac{X Y\left[r_S'-\left(r_P+r_S\right)\cot \theta  \right]}{(Z-i \Lambda )^2} \nonumber \\
& \,-i f_P  \frac{    r_P'+\left(r_P+r_S\right)\cot \theta}{Z-i \Lambda }. \label{vErefz}
\end{align}
For sake of clarity, in the formulas above we have omitted the subscript ``$r$'' from the coordinates $X,Y,Z$, and we have used the shorthand
\begin{align}
r_A := r_A^0(\theta) \qquad r_A' := \frac{\partial \, r_A^0}{\partial \, \theta}(\theta) \qquad r_A'' := \frac{\partial^{\,2} \, r_A^0}{\partial \, \theta^2}(\theta), \qquad (A \in \{P,S \}).
\end{align}
Moreover, as the variable $D$ appears always in the form $Z_r + D$, in the equations above with $Z$ we denoted $Z_r + D$, which amounts to a trivial re-definition of the origin of $K_r$.

It is easy to see that the expressions for the electric field obtained above take explicitly the form of a power series expansion in the parameter $\theta_0$ if we redefine the coordinates as
\begin{align}\label{rescale}
X = k_0 x_r = \frac{2}{\theta_0} \xi, \quad Y = k_0 y_r = \frac{2}{\theta_0} \eta, \quad Z = k_0(z_r + D) = \frac{2}{\theta_0^2} \zeta,
\end{align}
where $\xi = x_r/w_0$, $\eta = y_r/w_0$, and $\zeta  = (z_r + D/k_0)/L$. After this rescaling, Eq. (\ref{vEref}) takes the form of a power series:
\begin{align}\label{vEref3}
 \mathbf{E}^R (\mathbf{r})
 &  \simeq   \mathbf{E}^R_0 (\mathbf{r}) + \theta_0 \mathbf{E}^R_1 (\mathbf{r})+ \theta_0^2 \mathbf{E}^R_2 (\mathbf{r}),
\end{align}
where we have omitted an irrelevant overall multiplicative factor $\theta_0^2$. Finally, from Eq. (\ref{vEref3}) the field intensity may be straightforwardly calculated as
\begin{align}\label{Intensy}
I (\mathbf{r})
 &  \simeq   \abs{\mathbf{E}^R_0}^2  + \theta_0 \left( \mathbf{E}^R_0 \cdot {\mathbf{E}^R_1}^* + \mathrm{c.c.} \right) + \frac{1}{2} \theta_0^2 \left( \abs{\mathbf{E}^R_1}^2 + 2 \mathbf{E}^R_0 \cdot {\mathbf{E}^R_2}^* + \mathrm{c.c.} \right)+ \mathcal{O}(\theta_0)^3,
\end{align}
where ``$ \mathrm{c.c.}$'' stands for \emph{complex conjugate}. The explicit expression for $I (\mathbf{r})$ is quite cumbersome and it will not be reported here.

At this point, we have all the ingredients to calculate Eq. (\ref{Xmedio}) that gives
\begin{align}
\left\langle X_r \right\rangle
 = &  - \frac{Z}{\Lambda} \frac{\abs{f_P}^2 r_P\, r_P' + \abs{f_S}^2 r_S \, r_S' }{\abs{f_P}^2 (r_P^2 + \epsilon_P) +  \abs{f_S}^2 (r_S^2 + \epsilon_S)}, \label{nat40}
\end{align}
 where $\epsilon_P$ represents the contribution of second order terms in the Taylor expansion and it is defined by
\begin{align}\label{nat50}
 \epsilon_P  =  \frac{1}{2 \Lambda}\left[{r_P'} + r_P \, r_P'' - r_S^2+ r_P \, r_P' \cot\theta + (r_P^2 - r_S^2) \csc^2 \theta    \right].
\end{align}
Here $\epsilon_S$ is obtained from $\epsilon_P$ by interchanging the indices $P$ and $S$. Note that for a $TM$-polarized beam at Brewster incidence $r_P=0$ and $f_S=0$, and the denominator of Eq. (\ref{nat50})  remains non zero only thanks to $\epsilon_P$.

Equation (\ref{nat40}) shows that the distance from the beam center to the reference axis $z_r$ grows linearly with $Z$ as $\left\langle X_r \right\rangle =  Z \Theta  $, thus defining unambiguously the  \emph{angular shift} of the beam equal to $\Theta = \partial \left\langle X_r \right\rangle/ \partial Z $ \cite{NotePorras}. This definition is purely analytical and therefore, contrarily to the geometric one adopted by several authors \cite{ChanANDTamir,Greffet92}, it is always valid, even in the case of strong deformation or splitting of the reflected beam.
In our experimental setup, beam reflection occurs at the front surface of a BK7 prism with refractive index $n = 1.51031$ at $826$ nm, which corresponds to a Brewster angle $\theta_B =\arctan n= 56.491^\circ$. 
For a \textsl{TM}-polarized incident beam ($f_P=1$ and $f_S=0$), Eq. (\ref{nat40}) becomes
\begin{align}\label{nat60}
\Theta = -\frac{1}{\Lambda}\frac{r_P r_P'}{r_P^2 + \epsilon_P},
\end{align}
which shows that $\Theta = 0$ at $\theta_B$ where $r_P=0$ and $\epsilon_P \neq 0$. However, since $\epsilon_P \propto 1/\Lambda $,  and   $r_P \propto -(\theta - \theta_B)$ nearby $\theta_B$,  then in this region Eq. (\ref{nat60}) reduces approximately to
\begin{align}\label{nat70}
\Theta \cong \frac{\theta - \theta_B}{\Lambda(\theta - \theta_B)^2 + \alpha},
\end{align}
where
\begin{align}\label{nat71}
\alpha = \frac{1}{2} + \frac{2 n^4 }{\left(1+n^2\right)^4}.
\end{align}
Since $\alpha$ does not depend on $\Lambda$, it is easy to see
  from Eq. (\ref{nat70}) that if we put $\Theta = f(x)$, with
  $x = \theta - \theta_B$, then the following scaling property holds:
\begin{align}\label{nat72}
\sqrt{\Lambda} f(x/\sqrt{\Lambda}) = \frac{x}{x^2 + \alpha}.
\end{align}
 Thus, there exists an angle $\theta_M = \theta_B + \sqrt{\alpha/\Lambda}$, ($\theta_m = \theta_B -\sqrt{\alpha/\Lambda}$) close to $\theta_B$ where $\Theta$ reaches a maximum (a minimum) approximately equal to $1/(2 \sqrt{\alpha \Lambda})$ [$-1/(2 \sqrt{\alpha \Lambda})$].
Since $\Lambda = (k_0 w_0)^2/2 = 2/\theta_0^2$, where $\theta_0$ is the angular spread of the incident beam \cite{MandelBook}, then the maximum angular displacement occurring at $\theta_M =  \theta_B+\theta_0 \sqrt{\alpha /2} \sim \theta_B+0.54 \, \theta_0$  will amount to $1/(2 \sqrt{\alpha \Lambda}) =  \theta_0 /\sqrt{8 \alpha } \sim 0.46 \, \theta_0 < \theta_0$. Moreover, for $|\theta - \theta_B| \gg \theta_0$, Eq. (\ref{nat70}) furnishes %
\begin{align}\label{nat80}
\frac{\Theta}{\theta_0} \sim \frac{\theta_0}{2(\theta - \theta_B)} \ll 1,
\end{align}
which is a signature of the \emph{sub-diffractive} nature of the phenomenon.
In Fig. 2  approximate expression (\ref{nat70})  is compared with the exact result (\ref{nat60}) for a beam waist $w_0 = 30$ $\mu$m. The agreement between the two curves is very good and we have verified that it increases for increasing $w_0$.
%
%
  \begin{flushleft}
  \begin{figure}[!ht]
  \centerline{\includegraphics[width=13truecm]{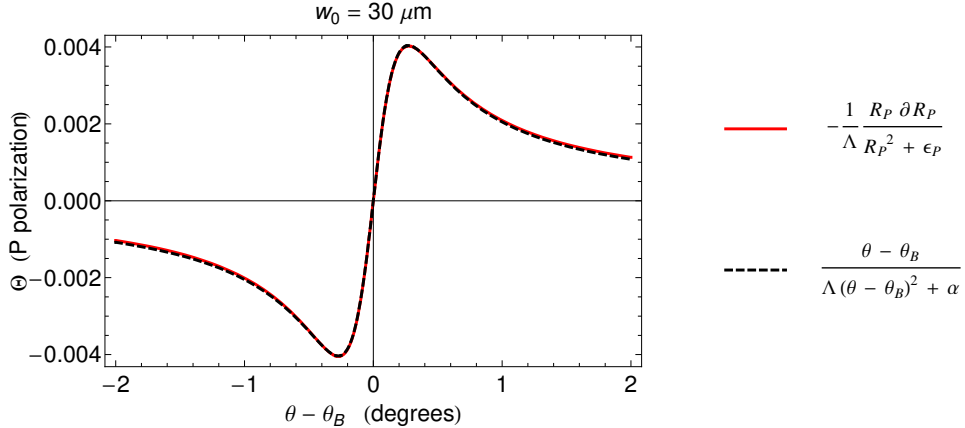}}
  \caption{Geometry of beam reflection at the air-medium interface.}
  \end{figure}
  \end{flushleft}
%
%
\section{Alternative route}
Both Goos-H\"{a}nchen and Imbert-Fedorov shifts can be calculated in an alternative manner that displays the both spatial and angular characters of the shifts. The starting point is Eq. (\ref{Eref2}) we rewrite here as
\begin{align}\label{Eref3}
 \mathbf{E}^R (\mathbf{r})
 &  = \iint\limits_{-\infty}^{\hphantom{xxii}\infty}   \bm{E} (U,V)e^{ i W Z} e^{i \left( - U X_r + V Y_r \right)} \; \di U \, \di V,
\end{align}
where $Z = Z_r + D$ and 
\begin{align}\label{Eref4}
  \bm{E} (U,V)
 &  =\sum_{\lambda=1}^2  \hve_\lambda(\widetilde{\vk}) r_\lambda(U,V) E_\lambda(U,V).
\end{align}
After a straightforward calculation, it is not difficult to prove the validity of the following formulas:
\begin{align}\label{deno}
\iint\limits_{-\infty}^{\hphantom{xxii}\infty} \abs{ \mathbf{E}^R (X,Y,Z)}^2 \; \di X \, \di Y
 &  = \iint\limits_{-\infty}^{\hphantom{xxii}\infty}   \abs{ \bm{E}(U,V)}^2 \;  \di U \, \di V,
\end{align}
\begin{align}\label{numX}
\iint\limits_{-\infty}^{\hphantom{xxii}\infty} X \abs{ \mathbf{E}^R (X,Y,Z)}^2 \; \di X \, \di Y
 &  = -i \iint\limits_{-\infty}^{\hphantom{xxii}\infty}   \frac{\partial \bm{E}}{\partial U} \cdot  \bm{E}^* \;  \di U \, \di V - Z \iint\limits_{-\infty}^{\hphantom{xxii}\infty}   \frac{U}{W}  \abs{ \bm{E}}^2  \;  \di U \, \di V,
\end{align}
\begin{align}\label{numY}
\iint\limits_{-\infty}^{\hphantom{xxii}\infty} Y \abs{ \mathbf{E}^R (X,Y,Z)}^2 \; \di X \, \di Y
 &  = i \iint\limits_{-\infty}^{\hphantom{xxii}\infty}   \frac{\partial \bm{E}}{\partial V} \cdot  \bm{E}^* \;  \di U \, \di V + Z \iint\limits_{-\infty}^{\hphantom{xxii}\infty}   \frac{V}{W}  \abs{ \bm{E}}^2  \;  \di U \, \di V.
\end{align}
Thus, we easily obtain
\begin{align}\label{GH}
\left\langle X \right\rangle
 &  = \frac{\displaystyle{-i \iint\limits_{-\infty}^{\hphantom{xxii}\infty}   \frac{\partial \bm{E}}{\partial U} \cdot  \bm{E}^* \;  \di U \, \di V}}{\displaystyle{\iint\limits_{-\infty}^{\hphantom{xxii}\infty}   \abs{ \bm{E}(U,V)}^2 \;  \di U \, \di V} }- Z \frac{\displaystyle{\iint\limits_{-\infty}^{\hphantom{xxii}\infty}   \frac{U}{W}  \abs{ \bm{E}}^2  \;  \di U \, \di V}}{\displaystyle{\iint\limits_{-\infty}^{\hphantom{xxii}\infty}   \abs{ \bm{E}(U,V)}^2 \;  \di U \, \di V}},
\end{align}
\begin{align}\label{IF}
\left\langle Y \right\rangle
 &  = \frac{\displaystyle{i \iint\limits_{-\infty}^{\hphantom{xxii}\infty}   \frac{\partial \bm{E}}{\partial V} \cdot  \bm{E}^* \;  \di U \, \di V}}{\displaystyle{\iint\limits_{-\infty}^{\hphantom{xxii}\infty}   \abs{ \bm{E}(U,V)}^2 \;  \di U \, \di V} }+ Z \frac{\displaystyle{\iint\limits_{-\infty}^{\hphantom{xxii}\infty}   \frac{V}{W}  \abs{ \bm{E}}^2  \;  \di U \, \di V}}{\displaystyle{\iint\limits_{-\infty}^{\hphantom{xxii}\infty}   \abs{ \bm{E}(U,V)}^2 \;  \di U \, \di V}}.
\end{align}
The equations above show clearly the spatial and the angular contributions to the shifts. The angular part is the part proportional to $Z$. It is interesting to note that the $Z$-dependence is \emph{strictly} linear, as these equations are exact. Moreover, if we remember that $U = k_{x_i}$ and $W = k_{z_i}$, then it is obvious that 
\begin{align}\label{nota}
\frac{U}{W} = \tan \theta_x, \qquad \text{and} \qquad \frac{V}{W} = \tan \theta_y.
\end{align}
From Eq. (\ref{nota}) it immediately follows that 
\begin{align}\label{GH2}
\frac{\partial \left\langle X \right\rangle}{\partial Z}
 &  = - \left\langle \tan \theta_x \right\rangle \simeq - \left\langle  \theta_x \right\rangle,
\end{align}
\begin{align}\label{IF2}
\frac{\partial \left\langle Y \right\rangle}{\partial Z}
 &  =  \left\langle \tan \theta_y \right\rangle \simeq  \left\langle  \theta_y \right\rangle.
\end{align}
%

%

\begin{thebibliography}{13}
\bibitem[{Not()}]{NotePorras}
\bibinfo{note}{This result is in agreement with Ref. \cite{Porras97}, where the
  author found the same linear dependence by calculating the moments of the
  Wigner distribution function of the reflected beam.}
\end{thebibliography}
%

\end{document}